\documentclass[11pt,a4paper]{article}
\usepackage{jheppubmcc}

\usepackage{latexsym}
\usepackage{amsfonts}
\usepackage{amssymb}
\usepackage{slashed}
\usepackage{feynmp}

\usepackage{amsmath}    
\usepackage{graphicx}   
\usepackage{verbatim}   
\usepackage{color}      
\usepackage{subfigure}  
\usepackage{hyperref}   
\usepackage{acronym}

\usepackage{bbm}
\usepackage{verbatim}   
\usepackage{subfigure}  
\usepackage{acronym}

\begin{document}

\title{Asymmetric Dark Matter via Spontaneous Co-Genesis}

\author[a]{John March-Russell,}
\emailAdd{jmr@thphys.ox.ac.uk}

\author[a]{Matthew McCullough}
\emailAdd{mccull@thphys.ox.ac.uk}

\affiliation[a]{Rudolf Peierls Centre for Theoretical Physics,
University of Oxford\\
1 Keble Road, Oxford,
OX1 3NP, UK}

\date{\today}

\preprintmcc{OUTP-11-42P}

\abstract{We investigate, in the context of asymmetric dark matter (DM), a new mechanism of spontaneous co-genesis of linked DM and baryon asymmetries, explaining the observed relation between the baryon and DM densities, $\Omega_{DM}/\Omega_B\simeq 5$.    The co-genesis mechanism requires a light scalar field, $\phi$, with mass below 5 eV which couples derivatively to DM, much like a `dark axion'.  The field $\phi$ can itself provide a final state into which the residual symmetric DM component can annihilate away.
}

\maketitle

\section{Introduction}\label{intro}

Usually it is assumed that the baryon and dark matter (DM) densities we observe in the universe today are generated by independent processes.  In particular, the baryon density is entirely determined by a CP-violating asymmetry between baryon and anti-baryon densities in the early universe, while the standard assumption has been that the DM density does not depend on any corresponding asymmetry and is instead determined by thermal freeze-out \cite{Zeldovich1,Zeldovich2,Chiu:1966kg} or thermal freeze-in \cite{Hall:2009bx}.  Since the genesis mechanisms for DM and baryons are thus decoupled, the DM-to-baryon ratio, $\Omega_{DM}/\Omega_B$, could in principle lie far away from the observed close coincidence, $\Omega_{DM}/\Omega_B \simeq 5$  \cite{Nakamura:2010zzi}.

An alternative picture is that the DM itself possesses a particle-antiparticle asymmetry, linked in some way to the
baryon asymmetry, which determines (at least the dominant component) of the DM density, thus explaining the observational fact
that $\Omega_{DM}/\Omega_B \simeq 5$  This proposal goes under the name Asymmetric Dark Matter (ADM) \cite{Hut:1979xw,Nussinov:1985xr,Gelmini:1986zz, Chivukula:1989qb,Barr:1990ca,Kaplan:1991ah,Thomas:1995ze,Hooper:2004dc,Kitano:2004sv,Agashe:2004bm,Cosme:2005sb,Farrar:2005zd,Suematsu:2005kp,Tytgat:2006wy,Banks:2006xr,Khlopov:2008ty,Kitano:2008tk,Kaplan:2009ag,Kohri:2009yn,Kribs:2009fy,Cohen:2009fz,An:2009vq,Khlopov:2010pq,Cohen:2010kn,Shelton:2010ta,Davoudiasl:2010am,Haba:2010bm,Buckley:2010ui,Gu:2010ft,Blennow:2010qp,Hall:2010jx,Dutta:2010va,Falkowski:2011xh,Heckman:2011sw,Graesser:2011wi,Frandsen:2011kt,Buckley:2011kk,Hook:2011tk,Cheung:2011if,DelNobile:2011je}.  Recently there has been a burst of activity considering the possibility of ADM in which the baryon and DM densities are determined by such linked asymmetries, though no standard picture has yet emerged.

Our aim in this paper is to demonstrate that there exists a new mechanism of {\it spontaneous co-genesis} of linked baryon and dark matter asymmetries.  We find that this mechanism for the generation of the asymmetries possesses a number of attractive features
compared to previous approaches.  As a prelude to our argument we start with a brief outline of the primary idea.

As is well known \cite{Sakharov:1967dj} the generation of a baryon-number asymmetry in the early Universe requires both CP violation and baryon-number violation.  In addition, if the underlying theory is CPT invariant, so that particle and anti-particle masses and energy eigenvalues are equal, an asymmetry requires a departure from thermal equilibrium.   However, as noted some time ago by Cohen and Kaplan \cite{Cohen:1987vi,Cohen:1988kt}, the expansion of the universe spontaneously violates both T and CPT, allowing, in principle, the generation of a baryon asymmetry in thermal equilibrium if there are sufficiently large differences between particle and anti-particle energy eigenstates.   More precisely, baryon-number violating scattering and decay processes can be in equilibrium with rates greater than the Hubble expansion rate, but, nevertheless, different thermal distributions for baryons and anti-baryons can occur if the expansion leads to a background `potential' biasing particle versus antiparticle densities in the thermal bath.  CP-violating phases leading to differences in scattering or partial decay rates between particles and anti-particles are not necessary, the spontaneous T-violation of the background being sufficient. 

In the context of baryogenesis, there exist a number of detailed implementations of this mechanism, usually dubbed `Spontaneous Baryogenesis' \cite{Cohen:1987vi,Cohen:1988kt,Cohen:1991iu,Abel:1992za,Cohen:1994ss,Dolgov:1994zq,Dolgov:1996qq,Bertolami:1996cq,Li:2001st,Yamaguchi:2002vw,Brandenberger:2003kc,Takahashi:2003db,Alberghi:2003ws,Carmona:2004xc,Carroll:2005dj,Barenboim:2007tu}.   It is natural to ask whether an adaption of this mechanism can lead to a mechanism of spontaneous co-genesis of both the baryon asymmetry and a dark matter asymmetry.   

One significant issue with previous implementations of the spontaneous genesis idea is that they require a new light degree
of freedom, usually a neutral scalar field $\phi$, derivatively coupled to baryon- or lepton-number carrying states, and with a time-dependent vev in the early universe.   Moreover, this time-dependent vev must not be in its oscillating phase during the epoch of spontaneous genesis, or the resulting asymmetry is severely suppressed \footnote{This suppression has been discussed by Dolgov and collaborators  \cite{Dolgov:1994zq,Dolgov:1996qq}, and invalidates some of the implementations of spontaneous baryogenesis studied in the literature.}.  Together with laboratory and cosmological/astrophysical constraints, these requirements severely limit the utility of the spontaneous genesis mechanism.   Our implementation, however, automatically solves this difficulty as the required light time-dependent field is now coupled to the DM, generating the asymmetry in the dark sector, rather than SM states, greatly ameliorating the constraints.  In addition, an unexpected and highly appealing consequence of our implementation of spontaneous
co-genesis, is the fact that the new field $\phi$, can naturally solve a generic problem of models of ADM, namely the efficient elimination of the symmetric part of the DM density, so that the final DM density is determined by the asymmetry alone.   We view this as a very attractive added benefit of our mechanism of spontaneous co-genesis.

In Section~\ref{genesis} we introduce the basic mechanism of spontaneous genesis in the DM sector.   In
Section~\ref{sharing} we discuss how the resulting DM asymmetry is shared with (equivalently, partially transferred to) the visible sector
via suitable `sharing interactions' between the dark and visible sectors.  We argue that along with the standard sharing paradigm where a fixed DM asymmetry is shared, there is a new regime where the DM asymmetry continues to evolve after the
sharing interactions drop out of equilibrium, allowing very different DM masses and interactions compared to the usual ADM case.
In Section~\ref{cosmologyphi} we discuss the cosmology of the new degree of freedom $\phi$ whose time-dependence drives the
spontaneous co-genesis mechanism.  We show, in particular, that additional interactions between $\phi$ and the DM can naturally lead to sufficient elimination of the symmetric component of the DM density.

\section{Introduction to spontaneous matter genesis}\label{genesis}

The CPT- and T-violation necessary for spontaneous co-genesis can arise dynamically, either through the Lorentz-violating vev of some vector field, or more simply through the time-dependence of a scalar field, $\phi$, which, for simplicity, we here take to be a neutral scalar with derivative couplings
to DM states.  If, after inflation, the scalar field does not lie at the minimum of its low-temperature potential then it will evolve towards this minimum as the Universe cools.  However, if the mass satisfies $m_\phi \lesssim 3 H$, where $H$ is the Hubble parameter at a given temperature, then the evolution of the field will be damped.  Assuming a spatially homogeneous field, as one would expect after inflation, then as the field evolves a homogeneous Lorentz-violating vev arises; $\partial_\mu \phi = \{\dot{\phi},\bold{0}\}$.  Thus if $\phi$ is derivatively coupled to some current, which we call $X$-number current, as
\begin{equation}
\mathcal{L} \supset \frac{\partial_\mu \phi}{f} J_X^\mu ~~,
\label{eq:current}
\end{equation}
where $f$ is a decay constant, the slow evolution of $\phi$ leads to an effective background potential for $X$-number density.  If there are $X$-number violating processes occurring at a rate $\Gamma_{\not{X}} > H$, a non-zero $X$-number is generated in thermal equilibrium.  The $X$-number violating processes are necessary, as although the background potential makes it energetically favorable to have a particle asymmetry this asymmetry can only develop if there are interactions which violate $X$-number.\footnote{One obvious candidate for $X$-number violation in thermal equilibrium is through `dark sphalerons'.  Alternatively one could allow for explicit non-renormalizable $X$-number violating operators arising due to physics in the UV.}

Although this mechanism has been previously considered in the context of baryogenesis where $X=B$, or $L$,  it could be responsible for the generation of an asymmetry in any class of particles charged under a continuous global $U(1)_X$ symmetry, where $X$ simply stands for an unknown symmetry.

In this work we propose a novel application of this mechanism whereby ADM is generated by the spontaneous genesis mechanism.  This asymmetry can be simultaneously or subsequently shared with the visible sector, leading to a connection between the baryon asymmetry and the dark matter asymmetry.  We posit a dark sector which exhibits a global $U(1)_X$ symmetry at low temperatures.\footnote{We expect that in a
theory including quantum gravitational effects all continuous global symmetries are violated, leading to the ultimate decay of  baryons and/or DM.  Since, however, such violation occurs through either higher-dimension operators suppressed by at least $M_{GUT}$, or through terms which are non-perturbatively small, the resulting lifetimes can be easily much greater than the Hubble time $1/H_0$.  As discussed in Section~\ref{sharing}, in this paper we will assume that there is an exact discrete symmetry - either a $\mathcal{Z}_2$ $X$-parity, or in the SUSY case $R$-parity, which stabilizes the DM.}  We also assume that the scalar $\phi$ is coupled to the $X$-current, as opposed to the baryon-number current, leading to an effective interaction
\begin{equation}
\mathcal{L} \supset \frac{\partial_\mu \phi}{f} J_X^\mu ~~ \Rightarrow ~~ U_X(T) (n_{X} - n_{\overline{X}})  ~~,
\label{eq:currentX}
\end{equation}
where $U_X (T) = \dot{\phi} (T)/f$ is the background potential for $X$-number.  Then, if the dark sector also exhibits $X$-number violating interactions, which freeze-out at a temperature $T_X$, this leads to an $X$-number asymmetry given by
\begin{equation}
X(T,U_X)  = \frac{T^3}{6} \frac{U_X}{T} g_X k\left(M_X/T,\pm1\right)  ~~,
\label{eq:asymm}
\end{equation}
where $g_X$ is the number of degrees of freedom of $X$, and the function $k (x,\pm1)$ is defined by
\begin{equation}
k (x,\pm1)  = \frac{6}{\pi^2}  \int^{\infty}_{x} \frac{\sqrt{y^2-x^2}}{(e^y\pm1)^2} y e^y dy ~~,
\label{eq:k}
\end{equation}
for fermions and bosons respectively.  For fermions in the relativistic and non-relativistic limits analytic forms for $k (x,+1)$ are
\begin{equation}
k \left (x,+1 \right ) \simeq  \left\{ \begin{array}{cl}
 1 & \mbox{ $(x \ll 1)$} \\
  12 \left ( \frac{x}{2 \pi } \right)^{3/2} e^{-x} & \mbox{ $(x \gg 1)$}~~.
       \end{array} \right .
\label{eq:k2}
\end{equation}

We will often wish to normalize the particle asymmetry given in eq.(\ref{eq:asymm}) by the entropy density at a given temperature in order to consider a dimensionless quantity which is not diluted by expansion
\begin{equation}
N_X (T) = \frac{X(T,U_X)}{s (T)}  ~~.
\label{eq:asymmtos}
\end{equation}

We can translate the required properties of the background potential into those of the rolling scalar field if we make the additional assumption that the evolution of $\phi$ is damped throughout the generation of $X$-number, in other words, $T_X$ is greater than the temperature at which $m_\phi \sim 3 H$.\footnote{For a discussion of the oscillating case see e.g.\ \cite{Dolgov:1996qq}.  Note that in the oscillating stage the asymmetry is parametrically smaller than in the damped case.  In this paper we will only be concerned with the situation where $\phi$ is non-oscillating.}  From the equation of motion, under the assumption of damping, we have
\begin{equation}
\dot{\phi} \simeq \frac{1}{3 H} \frac{dV_T(\phi)}{d\phi} \simeq \frac{M_P m_\phi^2 \phi_0}{5 g_\star^{1/2} (T) T^2}  ~~,
\label{eq:eqnmotion}
\end{equation}
where $V_T(\phi)$ is the thermal scalar potential, $M_P$ is the Planck mass, $m_\phi$ is the mass of the rolling scalar, $\phi_0$ is the vacuum expectation value of the scalar and $g_\star (T)$ is the effective number of relativistic degrees of freedom.  In the second equality in eq.(\ref{eq:eqnmotion}) we have made the approximation that thermal effects are subdominant and the potential can be well described by an effective mass term.

We impose the further constraint that the scalar motion remains non-oscillatory down to the temperature, $T_X$, at which $X$-number violation freezes out, thus we require that
\begin{eqnarray}
m_\phi & < & 1.66 g_\star^{1/2} (T_X) \frac{T_X^2}{M_P}  ~~.
\label{eq:scalardamping}
\end{eqnarray}
To encode this constraint we thus parameterize the scalar mass as
\begin{eqnarray}
m_\phi & = & \alpha \times 1.66 g_\star^{1/2} (T_X) \frac{T_X^2}{M_P}  \\
& \simeq & 2.1 \times 10^{-3} \alpha \left( \frac{T_X}{1 \text{ TeV}} \right)^2 \text{ eV}  ~~,
\label{eq:scalardamping2}
\end{eqnarray}
where, in the second line, for definiteness we have assumed a supersymmetric model with $g_\star \sim 250$.

By combining eq.(\ref{eq:asymmtos}) with eq.(\ref{eq:scalardamping2}) we find that
\begin{eqnarray}
\Omega_X h^2 (T_{now})=  5.9 \times 10^7  \cdot \frac{g_\star^{1/2} (T_X)}{g_{\star S} (T_X)}  \cdot \left[ \alpha^2 \frac{\phi_0}{f} \right] \cdot  \left[g_X \frac{M_X}{M_N} \frac{T_X}{M_P}\right] k \left (\frac{M_X}{T_X},+1 \right ) ~~,
\label{eq:darkmatter}
\end{eqnarray}
where $M_N$ is the mass of a nucleon and we have made the assumption that the dark matter particle is fermionic.  We have also assumed that interactions within the dark sector allow the efficient annihilation of the symmetric component of $X$-density to light, or massless states, such that the final DM density is determined purely by the asymmetry.  In Section~\ref{cosmologyphi} we return to this issue, and show that the symmetric component of the DM can annihilate away into $\phi$.

\begin{figure}[t]
\centering
\includegraphics[width=2.6in]{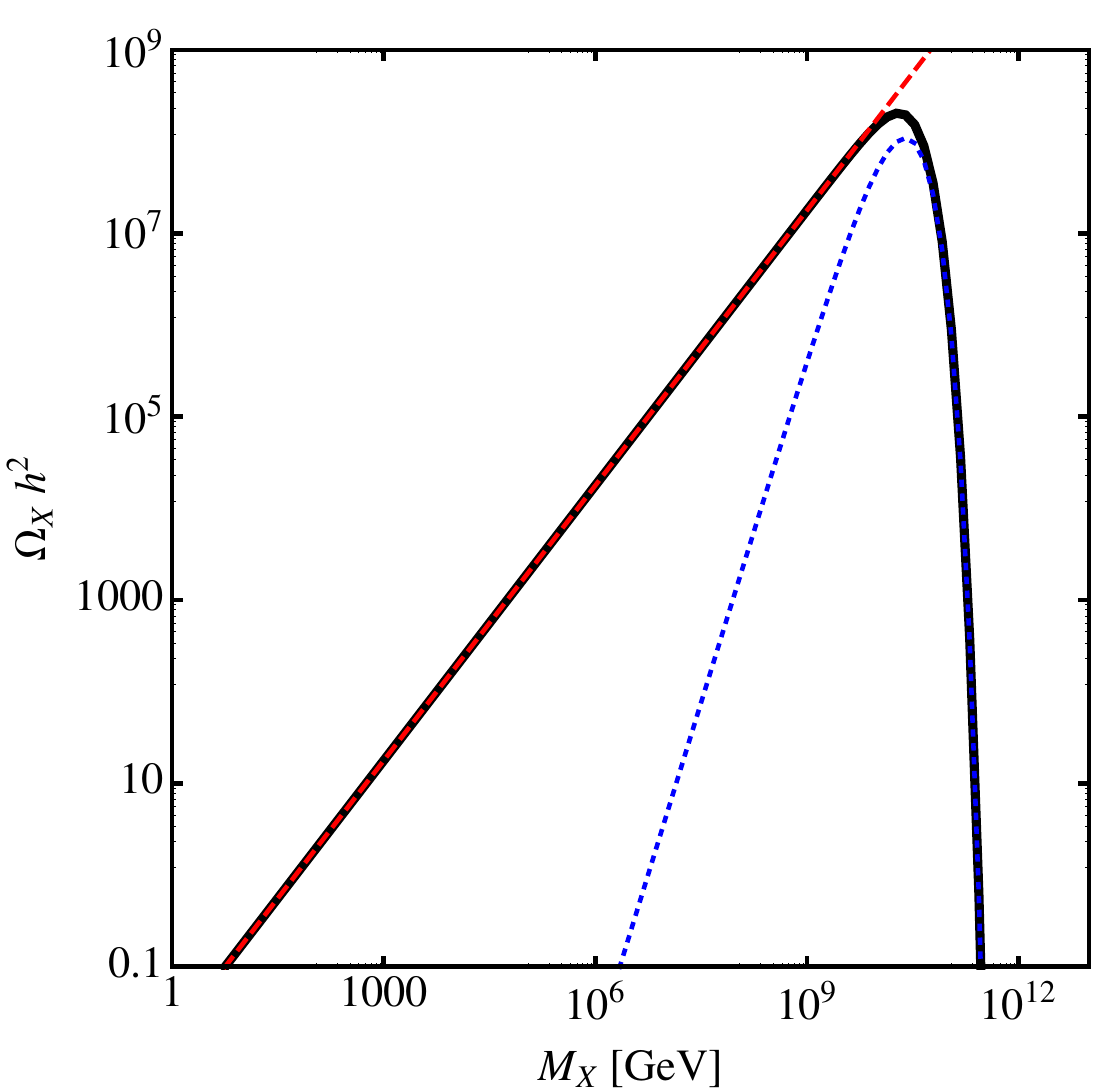} \hspace{0.4in} \includegraphics[width=2.8in]{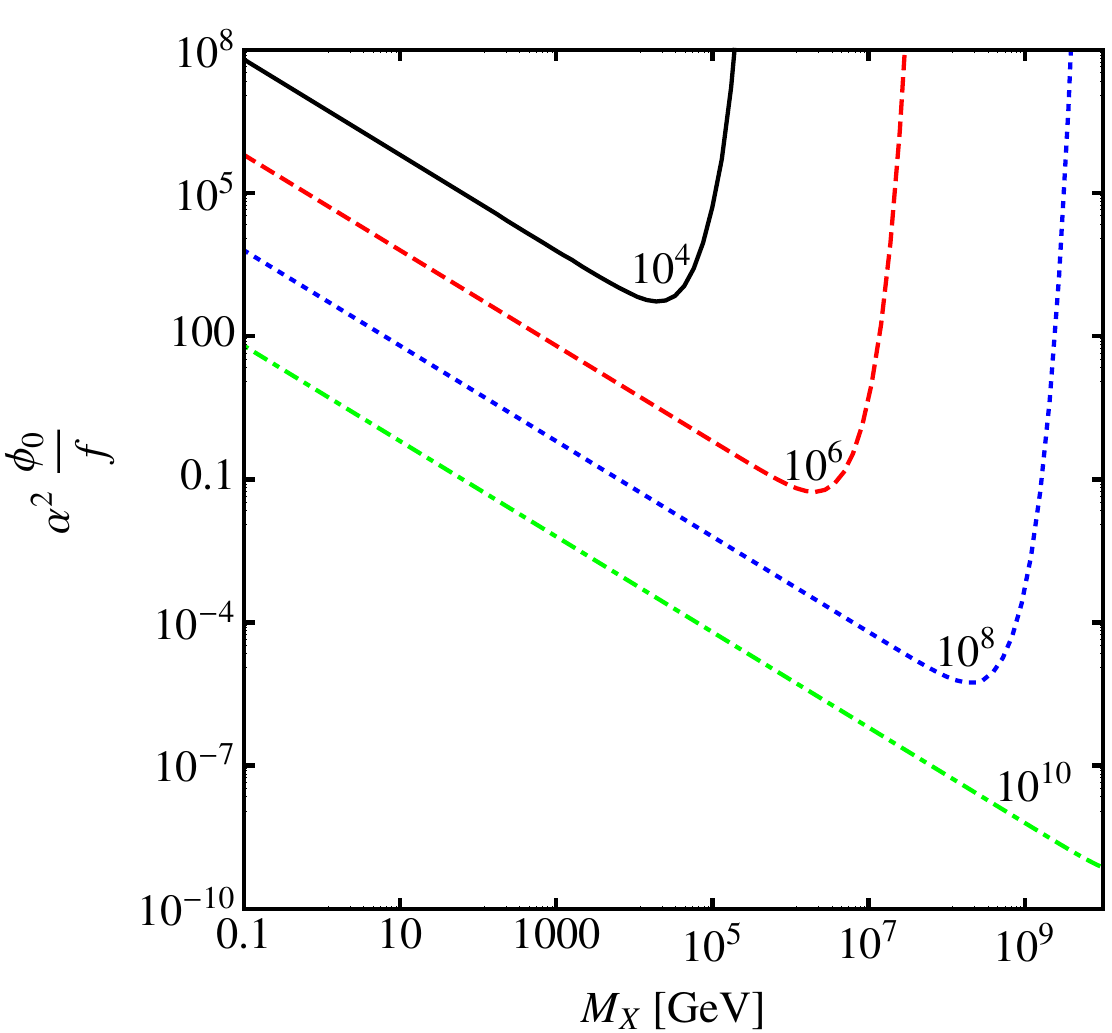}
\caption{Left panel:  The relic density of ADM energy density as a function of the DM mass for $\alpha^2 \phi_0/f = 1$ and $T_X = 10^{10}$ GeV.  The black line corresponds to the full solution calculated using eq.(\ref{eq:k}) and the red dashed (blue dotted) line corresponds to the relativistic (non-relativistic) approximation given in in eq.(\ref{eq:k2}).  Right panel:  The required values of the scalar field parameter combination $\alpha^2 \phi_0/f$ for varying DM mass at contours of fixed $T_X$, where the value of $T_X$ in GeV is labelled on each line.  As damped motion requires $\alpha < 1$, low values of $T_X\lesssim 10^5$ GeV require large values of $\phi_0/f$.}
\label{fig:RelicDensity}
\end{figure}

Eq.(\ref{eq:darkmatter}) is our master formula describing the current DM energy density.  The first term in brackets depends only on the properties of the rolling scalar field, where $\phi_0/f$ is the ratio of the initial vev of the scalar to the decay constant in the scalar-to-current coupling.  For an axion-like scalar this term will satisfy $\alpha^2 \phi_0/f \lesssim 1$, however if the field corresponds to some non-compact flat direction then this factor could be $\gg 1$.  The final quantities depend on the details of the dark sector; the DM mass $M_X$, $g_X$, and $T_X$.  In Figure~\ref{fig:RelicDensity} we plot the relic abundance as a function of the DM mass for a given choice of $T_X = 10^{10}$ GeV.  We also show contours in the $M_X - \alpha^2 \phi_0/f$ plane which generate the correct relic abundance for a given value of $T_X$.  

\section{Relation to the baryon asymmetry}\label{sharing}

\subsection{Sharing}

In order to connect a DM asymmetry to the observed baryon asymmetry via a sharing scenario it is necessary that, at high temperatures, $U(1)_{B-L} \times U(1)_X$ is broken down to a smaller group $U(1)_{(B-L+q X)}$, for some $q\neq 0$.  Specifically there must exist at least one operator, individually breaking $U(1)_{B-L}$ and $U(1)_X$, but conserving $U(1)_{B-L+qX}$, which mediates
sharing processes that are in thermal equilibrium.

As this operator could, in principle, mediate DM decay to baryons and leptons one must impose a DM stabilizing symmetry.  This could be a $\mathcal{Z}_2$, such as R-parity in a SUSY theory, an $X$-parity in a SUSY or non-SUSY theory, or a higher discrete symmetry.  For definiteness, in this paper we choose a model with an $X$-parity, $X\rightarrow -X$, as this allows us to consider masses $M_X \gg m_W$, as well as DM masses near, or below the weak scale.  We wish to emphasize that a SUSY model where R-parity is the DM stabilizing symmetry is also consistent with our spontaneous co-genesis mechanism.  This requires that the DM is the LSP.

For definiteness, we choose to focus on a supersymmetric model\footnote{It is equally possible to implement this mechanism in a non-supersymmetric framework.} described by the MSSM superpotential augmented by a dark matter Dirac mass term and a sharing operator
\begin{equation}
W_X = M_X \overline{X} X + \frac{1}{M_S^2} X^2 U^c D^c D^c ~~,
\label{eq:sharing}
\end{equation}
where $X$ is the dark matter chiral superfield carrying $X$-number $+1$, $U^c$ and $D^c$ are the usual MSSM right-handed quark superfields, and $M_S$ is the mass-scale of the sharing operator.

The sharing operator in eq.(\ref{eq:sharing}) can mediate $X$-number and baryon number violating interactions, but preserves $U(1)_{X +2 (B-L)}$.  We define $T_S$ as the temperature at which sharing interactions mediated by this operator freeze out.  For squark masses $m_{\tilde{q}} > 700$ GeV and a sharing scale $M_S > 1$ TeV, one finds that $T_S \gtrsim 70$ GeV.\footnote{In $W_X$ one can equivalently replace  $U^c D^c D^c$ by $L H_u$ if $T_S$ is greater than the freeze-out temperature, $T_{sph}$, for electroweak non-perturbative processes (sphalerons).  The reason for this is that while sphalerons are active the chemical potentials for SM particles satisfy $\mu_{u_L} + 2 \mu_{d_L}+ \mu_{\nu_{L}} = 0$, and the only continuous global symmetry in the MSSM is $(B-L)$.  As both $U^c D^c D^c$ and $L H_u$ have the same charge under this symmetry, these two operators lead to equivalent relative asymmetries.}

For the sharing operator of eq.(\ref{eq:sharing}) the chemical potentials for dark matter and right handed quarks must satisfy $2 \mu_X = \mu_{u_R} + 2 \mu_{d_R}$, and thus the $X$ and $B$ asymmetries are related at a given temperature.  Other relations between chemical potentials arise through MSSM Yukawa couplings, gauge interactions, and the requirement of charge neutrality of the Universe.  These relations are summarized in \cite{Harvey:1990qw,Chung:2008gv}.  

\begin{figure}[t]
\centering
\includegraphics[width=2.8in]{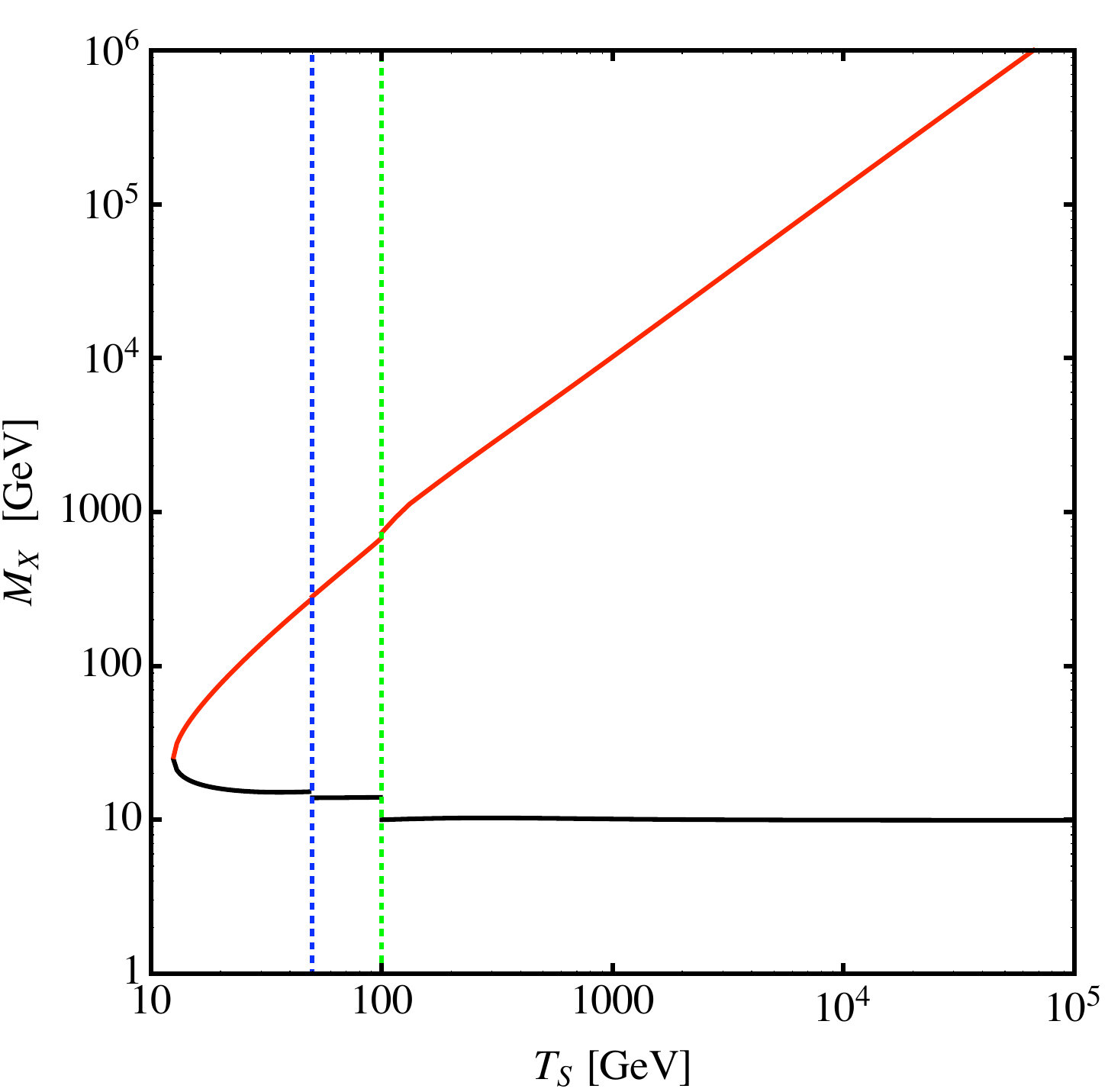}
\caption{Solutions, satisfying $\Omega_X/\Omega_B = 4.97$, for the ADM mass, $M_X$, as a function of the sharing freeze-out temperature, $T_S$.  We assume the sharing operator of eq.(\ref{eq:sharing}), and complete annihilation of the symmetric component of the DM density.  We illustrate a typical electroweak phase transition temperature by the vertical green dashed line, and a representative temperature, $T_{sph}$, at which sphalerons have become inactive, by the vertical blue dashed line.  For a given $T_S \gtrsim 20$ GeV there are two successful ADM solutions: One for $M_X \simeq 10$ GeV, where the DM is relativistic at $T_S$, while the other, non-relativistic solution has $M_X$ increasing with $T_S$ (as the DM density is Boltzmann suppressed in the non-relativistic regime).  This is the `sharing' paradigm.}
\label{fig:TXgtrTSmass}
\end{figure}

Employing the relations between chemical potentials it is possible to relate $X$-, $B$- and $L$-number asymmetries at a given temperature, resulting in relations of the form $X (T) = \gamma (T) B (T)$, where $\gamma (T)$ is a spectrum-dependent function which we have calculated following the methods in \cite{Harvey:1990qw,Chung:2008gv}.  If both the dark matter and baryons are relativistic at a given temperature then $\gamma (T) \sim \mathcal{O}(1)$ and a DM solution exists for $M_X \sim 10$ GeV.  However if the dark matter particles are non-relativistic, i.e.\ $T\ll M_X$, but some baryon number carrying state is still relativistic, then $\gamma (T)$ will be exponentially small and the correct DM density requires $M_X \sim 10 T_S$.  Both solutions are shown in Figure~\ref{fig:TXgtrTSmass}.\footnote{Complications arise if lepton-flavour violation is assumed to be out of thermal equilibrium, however as this is not central to the mechanism we are considering we assume this is not the case throughout.}  It should be noted that if, in the sharing operator in eq.(\ref{eq:sharing}), we replace $X^2$ with $X^n$ then the relativistic DM solution becomes $M_X \simeq 5 n$ GeV.  Hence, for superpotential operators linear in $X$, such as $X U^c D^c D^c$ or $X Q L D^c$, with the DM stabilized by $R$-parity, we would expect the DM mass to be closer to $5$ GeV.

\subsection{Spontaneous co-genesis}

When we combine the spontaneous genesis mechanism with the asymmetry sharing paradigm there now exist two distinct regimes.  If the spontaneous genesis completes before sharing has frozen out, i.e.\ $T_X>T_S$, then the DM mass is set by $T_S$.  Alternatively, there exists the possibility that sharing freezes out before spontaneous genesis in the dark sector has completed, so $T_S>T_X$.  In this case the dark matter asymmetry continues to evolve after the $B-L$ asymmetry has been set.  This scenario is of interest, as it leads to alterations to the standard relationship between the DM mass and $T_S$.

It may appear that by adding the sharing aspect to the spontaneous genesis mechanism we have also introduced an additional free parameter, namely $T_S$.  However this is not the case as we also have an additional constraint, given by the observed baryon asymmetry.  In total there are four parameters which govern the DM relic abundance, given by $T_X$, $T_S$, $M_X$, and the combination of scalar parameters, $\alpha^2 \phi_0/f$.  However there are two constraints: $\Omega_B h^2$ and $\Omega_{DM} h^2$.  Thus in total, this complete scenario only has two free parameters, which we choose to take as $T_X$ and $T_S$.\footnote{In order for our effective field theory description to remain consistent the temperatures must satisfy $T_X, T_S \lesssim f$.} 

\begin{figure}[t]
\centering
\includegraphics[width=3.4in]{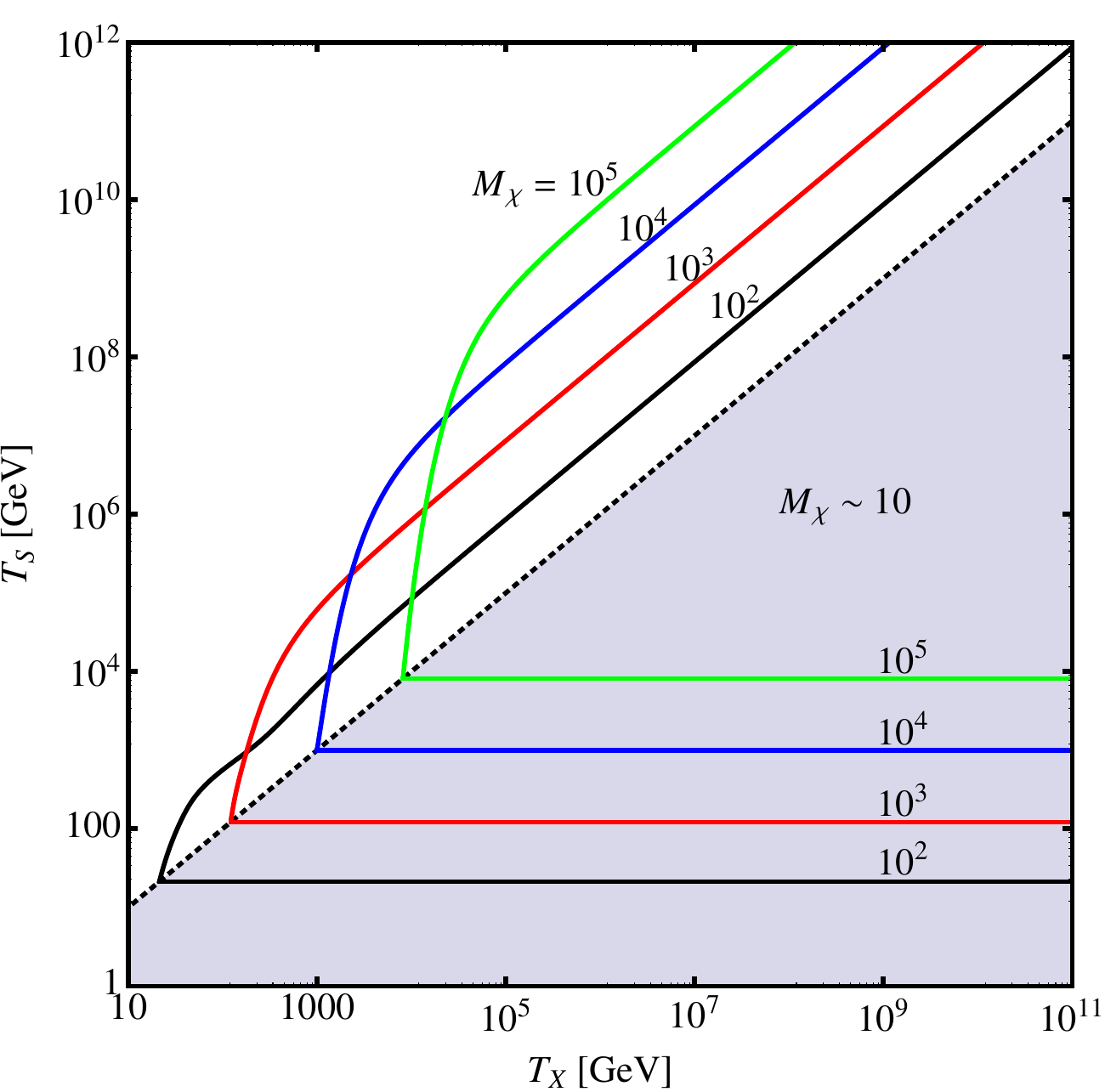}
\caption{Contours of constant $M_X$ in the $T_X$--$T_S$ plane corresponding to the generation of $\Omega_X/\Omega_B = 4.97$, and $\Omega_B h^2 =0.023$.    For $T_S < T_X$ there are two branches of solutions corresponding to the two branches shown in Figure~\ref{fig:TXgtrTSmass}: The first, relativistic solution
occurs when $M_X \sim 10$ GeV and fills the entire lower half plane (shaded region), while the second branch is shown by the horizontal contours
in the lower half plane labelled by $M_X$ in units of GeV.   Both solutions are independent of $T_X$ as in this case the sharing of the
asymmetry is determined after the total asymmetry has been frozen in.  On the other hand, for $T_X < T_S$ the DM asymmetry continues to evolve after sharing has ceased.  The resulting contours of constant $M_X$ corresponding to successful generation of $\Omega_X/\Omega_B$, and $\Omega_B h^2$ are
shown in the upper half plane, and the mass, $M_X$, now depends on both $T_X$ and $T_S$.  The portions of contours where $T_S \propto T_X$
apply for the relativistic case ($M_X\ll T_X$), while the remaining portions apply to the semi- and non-relativistic cases ($M_X\gtrsim T_X$).
The solution for $M_X \sim 10$ GeV lies along the line $T_S \simeq T_X$, and hence every point of this line corresponds to a solution when we continue to the $T_S < T_X$ corner, showing the continuity between solutions on either side of the line $T_X = T_S$. By allowing the DM asymmetry to evolve after sharing has ceased, a new set of solutions for a given DM mass and $T_S$ open up in the upper left half plane, in addition to the standard solutions in the lower half plane where $T_S<T_X$.}
\label{fig:TXTS}
\end{figure}

In Figure~\ref{fig:TXTS} we plot contours of constant DM mass in the $T_X$--$T_S$ plane which satisfy $\Omega_X/\Omega_B = 4.97$, and generate the observed
baryon asymmetry.  

One might wonder if it is still necessary to have $X$-number violation.  Without $X$-number violation there is a conserved $U(1)_{X+2 (B-L)}$ symmetry.  It can be imagined that, using the operator of eq.(\ref{eq:sharing}), one could create an equal and opposite asymmetry in $X$ and $2 (B-L)$, without ever violating $U(1)_{X+2 (B-L)}$.  However, by considering the various chemical potentials and conserved charges, it is straightforward to show that any asymmetry in $X$, $B$, or $L$-number is proportional to the total asymmetry in $X + 2 (B-L)$.\footnote{While sharing is active we have $2 \mu_X = \mu_{u_R} + 2 \mu_{d_R}$.  Rearranging this equation using additional relations between chemical potentials, found in \cite{Harvey:1990qw}, it can be shown that $\mu_X \propto \mu_B$, and an asymmetry in $X$ implies an asymmetry in $B$ of the same sign.  While sphalerons are active $\mu_{u_L} + 2 \mu_{d_L}  = - \mu_{\nu_{L}}$ and, using the previous relation, $\mu_X \propto -\mu_L$, hence an asymmetry in $X$ implies an asymmetry in $L$ of the opposite sign.  Consequently, if we create an asymmetry in $X$ then we must create a non-zero asymmetry in $X + 2 (B-L)$.}  Thus without violation of $U(1)_{X+2 (B-L)}$ it is impossible to create an asymmetry in DM or baryons.  As $B$ or $L$-violating operators are more tightly constrained, we choose to have the violation of $U(1)_{X+2 (B-L)}$ arising due to $X$-number violation in the dark sector, consistent with exact conservation of $X$-parity.

\section{Cosmology of $\phi$}\label{cosmologyphi}

We now consider the constraints on the scalar field parameters arising from production of the correct magnitude of baryon asymmetry, and limits on additional hot and cold dark matter components.  We also show that there exists a range of parameter space where
interactions between $\phi$ and the DM lead to efficient annihilation of the symmetric component of DM, leaving just the asymmetric component, as required for a complete theory of ADM.

\subsection{Baryon asymmetry and $\phi_0/f$}

From eq.(\ref{eq:darkmatter}) we see that the generated particle asymmetry is proportional to the combination of scalar parameters $\alpha^2 \phi_0/f$.  In order to generate a baryon-asymmetry of $N_B = 8.7 \times 10^{-11}$, we find the requirement 
\begin{equation}
\alpha^2 \frac{\phi_0}{f} \simeq \frac{10^{10} \text{GeV}}{\text{Max} [T_X,T_S]} ~~,
\label{eq:relations}
\end{equation}
where the exact relation depends on the details of the particle spectrum, but does not change by more than a factor of two when, eg,\ SUSY particles are included.   

Sharing is efficient during the generation of the baryon asymmetry, implying $U_B (T) \propto U_X (T)$.  At this temperature the majority of baryon-number carrying species are relativistic, hence, from eq.(\ref{eq:darkmatter}), $N_B (T) \propto T (\alpha^2 \phi_0/f)$.  As a result, if we wish to generate a specific baryon asymmetry for any $T$ we require that $\alpha^2 \phi_0/f \propto T^{-1}$.  For the case $T_X>T_S$ the total asymmetry is frozen in at $T_X$, whereas if $T_X<T_S$ the $(B-L)$ asymmetry is frozen in at $T_S$, explaining the form of eq.(\ref{eq:relations}).

Having $\text{Max}[T_X,T_S] \ll 10^{10}$ GeV requires $\phi_0 \gg f$, as $\alpha<1$.  For an axion-like scalar with a compact moduli space this is not possible, suggesting that the scalar should correspond to a non-compact flat direction such as might arise in supersymmetric models.

It should also be noted that we have assumed the simplest possible potential for $\phi$, with a single, temperature-independent mass term.  If this potential contained additional terms, or temperature dependence, such that at the time of spontaneous co-genesis $\frac{d V_T (\phi)}{d\phi} \gg m_\phi^2 \phi_0$ then it may be possible to achieve the required particle asymmetry, with $\phi_0 \sim f$ for $T_X \lesssim 10^{10}$ GeV.  The number of possibilities for such alterations, beyond the minimal model studied here, is large.

\subsection{Relic density of $\phi$ and bounds on $m_\phi$ and $T_X$}

There are constraints on the $\phi$ field parameters arising from the requirement that neither the energy density due to coherent
oscillations of $\phi$ nor the energy density of the thermally produced component of $\phi$ are too large.     

Regarding the thermal component, we assume that after inflation the Universe reheats to a temperature $T_I$.  If $\phi$ is in equilibrium with the dark sector and, through the sharing operator, the visible sector at early times, then the ratio of the number density to entropy density is roughly $1/g_\star (T_I)$, and the energy density at some later time is  $\rho_{therm} (T) \simeq m_\phi s (T)/g_\star (T_I)$
(we assume that $\phi$ is in thermal equilibrium at $T_I$ in order to set conservative constraints).

The calculation of the energy density in the coherent component is standard, and taking $g_\star (T_X) \simeq g_\star (\sqrt{\alpha} T_X) \simeq g_\star (T_I) \simeq 250$, for definiteness, then $T_{osc} \simeq \sqrt{\alpha} T_X$ and the relic energy density due to $\phi$ resulting from both production mechanisms is
\begin{eqnarray}
\Omega h^2_{osc} & \simeq & 1.2 \times 10^{-7} \sqrt{\alpha} \left ( \frac{T_X}{\text{TeV}} \right ) \left ( \frac{\phi_0}{10^{10} \text{GeV}} \right )^2 \\
\Omega h^2_{therm} & \simeq & 2.4 \times 10^{-6} \alpha \left ( \frac{T_X}{\text{TeV}} \right )^2 ~~.
\end{eqnarray}

The thermal component of $\phi$ behaves as hot dark matter, and thus we require that $\Omega h^2_{therm} \lesssim 0.007$, the current WMAP7 limit on relic energy density in neutrinos \cite{Komatsu:2010fb}.  Hence we require that $\sqrt{\alpha} T_X \lesssim 50$ TeV.  If we maximize the scalar mass by taking $\alpha =1$, then for $T_X \lesssim 50$ TeV (so satisfying the thermal bound), and $\phi_0 \lesssim 4 \times 10^{11}$ GeV (to satisfy the coherent bound with $\Omega h^2_{osc}<0.01$), the scalar field constitutes a subdominant component of the DM, hot or cold, leaving the ADM as the dominant component.  These inequalities then translate in to the requirement\footnote{This limit is on the current value of $m_\phi$, however from eq.(\ref{eq:eqnmotion}) we see that in order to generate a background potential $m_\phi$ must be non-zero at $T_X$.  Unlike the QCD axion mass, $m_\phi$ must be a UV-hard mass and be generated at high temperatures.  Such a non-perturbatively small, UV-hard, mass could arise if the shift symmetry $\phi\rightarrow\phi +\text{const}$ is broken by UV non-perturbative effects, such as string, or gravitational, instantons, or gauge instantons in a theory with a UV Landau pole \cite{Svrcek:2006yi,Arvanitaki:2009fg}.} 
\begin{equation}
m_\phi |_{\rm now} \lesssim 5~{\rm eV}.
\end{equation}
These bounds can be evaded if additional operators are included which enable $\phi$ to decay to lighter states.  For minimality we do not consider these additional operators here, however, we note that such decays are, in some parameter regions, strongly constrained.  

\subsection{Annihilation of symmetric DM component by $\overline{X} X \rightarrow \phi \phi$}

An appealing feature of the field $\phi$ is that, as well as generating a DM and baryon asymmetry, it can also enable the $X$-number symmetric component of DM density to annihilate away to light particles.  Whenever $X$-number is conserved the interaction in eq.(\ref{eq:current}) can be rearranged into a total derivative term, and thus doesn't allow for DM annihilation while $X$-number is conserved.  While $X$-number violation is efficient, and $\partial_\mu J^\mu_X \neq 0$, the interaction in eq.(\ref{eq:current}) may allow the symmetric component to annihilate away, however this depends on the source of $X$-number violation, and requires that the annihilation shuts off at the same time as the spontaneous genesis.

Alternatively, we can include additional fields and couplings to build a model which allows for efficient DM annihilation after $X$-number violation has ceased.  As a simple explicit example we consider fermionic DM and add an additional real scalar field $S$, with mass $M_S$, and couplings $\mathcal{L} \supset c_S S \overline{\psi}_X \psi_X + c_\phi S (\partial \phi)^2 / f$, where $c_S$ and $c_\phi$ are ${\cal O}(1)$ dimensionless coefficients.  These couplings respect the $U(1)_X$ symmetry, and preserve the shift symmetry of $\phi$, keeping it light.  This leads to a p-wave suppressed annihilation cross-section $X\bar{X}\rightarrow \phi\phi$
\begin{equation}
 \langle \sigma v \rangle \approx \frac{3 c_S^2 c_\phi^2}{16 \pi^2 f^2} \frac{M_X^4}{(4 M_X^2-M_S^2)^2+M_S^2 \Gamma_S^2} \frac{T}{M_X} ~~.
\end{equation}
Here we have made the mild assumption that $M_S>2m_\phi$.
 
By comparison with the results from \cite{Graesser:2011wi} we find that the ratio of symmetric to asymmetric DM densities is less than $10 \%$ if 
\begin{eqnarray}
f &\lesssim & 400 \frac{c_S c_\phi M_X^2}{\sqrt{(4 M_X^2-M_S^2)^2+M_S^2 \Gamma_S^2}} ~\text{GeV}~~,\\
&\lesssim & 100 c_S c_\phi ~\text{GeV}~~,
\end{eqnarray}
where in the second line we have taken $M_S < 2 M_X$ and (conservatively) assumed that the annihilation is not resonantly enhanced.
This constraint on $f$ is easily consistent with the other experimental and theoretical bounds on $f$.   Thus our model of ADM generation possesses a feature that we consider particularly attractive: $\phi$ both generates the DM and baryon asymmetry, and provides the final state into which the symmetric DM component annihilates away! This solves a significant problem of the ADM scenarios, as the operators allowing direct annihilation of the symmetric part of the DM to light SM states are strongly constrained by direct detection and collider bounds \cite{Kaplan:2009ag,Graesser:2011wi,Buckley:2011kk,Rajaraman:2011wf} (unless one arranges for annihilation solely to leptons).

\subsection{DM scattering and $\phi$}

One might also worry that the coupling $(\partial_\mu \phi) J^\mu_X/f$ for low $f$ could lead to unacceptable DM-DM scattering.  However the cross-section for such processes scales as $1/f^4$, and as $f\gtrsim 1$ GeV, the cross-section is well below current bounds \cite{Buckley:2009in}.

We also comment that light scalars derivatively coupled to DM, such as $\phi$, could lead to the novel process of enhanced stellar cooling through $\phi$-sstrahlung in DM-nucleon scattering.  These processes lead to bounds on $f$, in analogy with standard axion bounds.  These are, however, significantly weaker relative to standard axion bounds as these processes involve DM-nucleon (electron) scattering, and not nucleon-nucleon (electron) scattering.  Further, there is additional suppression due to the much lower density of DM compared to nucleons or electrons in a star.

\section{Summary}\label{conclusions}
We have described how, by derivatively coupling a light scalar, with a time-evolving expectation value, to the $X$-current, and allowing for $X$-number violating processes in the early Universe, it is possible to generate a DM asymmetry.  By utilizing `sharing' operators which allow for the transfer of particle asymmetries between the visible and dark sectors it is possible to simultaneously co-generate the observed baryon asymmetry and a DM particle asymmetry, providing a link between the two.  All of this occurs without the need for additional $CP$-violating parameters in either sector.  This is the spontaneous co-genesis mechanism.

The mechanism has a number of noteworthy aspects.  Most notable is the prediction of a light scalar with mass $m_\phi \lesssim 5$ eV.  In addition this scalar provides the attractive feature that it can automatically provide a final state for the annihilation of the symmetric DM component.

\acknowledgments
We gratefully thank Lawrence Hall, James Unwin and Christopher McCabe for stimulating discussions.    JMR is supported in part by both ERC Advanced Grant BSMOXFORD 228169, and a Royal Society Wolfson Merit Award, and MM and JMR both acknowledge support from EU ITN grant UNILHC 237920 (Unification in the LHC era).  MM thanks the STFC for support by a Postgraduate Studentship.

\bibliographystyle{JHEP}
\bibliography{SpontDarkrefs}

\providecommand{\href}[2]{#2}\begingroup\raggedright\begin{thebibliography}{10}

\bibitem{Zeldovich1}
Y.~B. Zel'dovich {\em Zh. Eksp. Teor. Fiz.} {\bf 48} (1965) 986.

\bibitem{Zeldovich2}
Y.~B. Zel'dovich, L.~B. Okun, and S.~B. Pikelner {\em Usp. Fiz. Nauk.} {\bf 84}
  (1965) 113.

\bibitem{Chiu:1966kg}
H.-Y. Chiu, {\it {Symmetry between particle and anti-particle populations in
  the universe}},  {\em Phys. Rev. Lett.} {\bf 17} (1966) 712.

\bibitem{Hall:2009bx}
L.~J. Hall, K.~Jedamzik, J.~March-Russell, and S.~M. West, {\it {Freeze-In
  Production of FIMP Dark Matter}},  {\em JHEP} {\bf 03} (2010) 080,
  [\href{http://xxx.lanl.gov/abs/0911.1120}{{\tt arXiv:0911.1120}}].

\bibitem{Nakamura:2010zzi}
{\bf Particle Data Group} Collaboration, K.~Nakamura {\em et.~al.}, {\it
  {Review of particle physics}},  {\em J. Phys.} {\bf G37} (2010) 075021.

\bibitem{Hut:1979xw}
P.~Hut and K.~A. Olive, {\it {A cosmological upper limit on the mass of heavy
  neutrinos}},  {\em Phys. Lett.} {\bf B87} (1979) 144--146.

\bibitem{Nussinov:1985xr}
S.~Nussinov, {\it {Technocosmology: could a technibaryon excess provide a
  'natural' missing mass candidate?}},  {\em Phys.Lett.} {\bf B165} (1985) 55.

\bibitem{Gelmini:1986zz}
G.~B. Gelmini, L.~J. Hall, and M.~J. Lin, {\it {What is the cosmion?}},  {\em
  Nucl. Phys.} {\bf B281} (1987) 726.

\bibitem{Chivukula:1989qb}
R.~Chivukula and T.~P. Walker, {\it {Technicolor cosmology}},  {\em Nucl.Phys.}
  {\bf B329} (1990) 445.

\bibitem{Barr:1990ca}
S.~M. Barr, R.~Chivukula, and E.~Farhi, {\it {Electroweak fermion number
  violation and the production of stable particles in the early universe}},
  {\em Phys.Lett.} {\bf B241} (1990) 387--391.

\bibitem{Kaplan:1991ah}
D.~B. Kaplan, {\it {A Single explanation for both the baryon and dark matter
  densities}},  {\em Phys.Rev.Lett.} {\bf 68} (1992) 741--743.

\bibitem{Thomas:1995ze}
S.~D. Thomas, {\it {Baryons and dark matter from the late decay of a
  supersymmetric condensate}},  {\em Phys.Lett.} {\bf B356} (1995) 256--263,
  [\href{http://xxx.lanl.gov/abs/hep-ph/9506274}{{\tt hep-ph/9506274}}].

\bibitem{Hooper:2004dc}
D.~Hooper, J.~March-Russell, and S.~M. West, {\it {Asymmetric sneutrino dark
  matter and the Omega(b) / Omega(DM) puzzle}},  {\em Phys.Lett.} {\bf B605}
  (2005) 228--236, [\href{http://xxx.lanl.gov/abs/hep-ph/0410114}{{\tt
  hep-ph/0410114}}].

\bibitem{Kitano:2004sv}
R.~Kitano and I.~Low, {\it {Dark matter from baryon asymmetry}},  {\em
  Phys.Rev.} {\bf D71} (2005) 023510,
  [\href{http://xxx.lanl.gov/abs/hep-ph/0411133}{{\tt hep-ph/0411133}}].

\bibitem{Agashe:2004bm}
K.~Agashe and G.~Servant, {\it {Baryon number in warped GUTs: Model building
  and (dark matter related) phenomenology}},  {\em JCAP} {\bf 0502} (2005) 002,
  [\href{http://xxx.lanl.gov/abs/hep-ph/0411254}{{\tt hep-ph/0411254}}].

\bibitem{Cosme:2005sb}
N.~Cosme, L.~Lopez~Honorez, and M.~H. Tytgat, {\it {Leptogenesis and dark
  matter related?}},  {\em Phys.Rev.} {\bf D72} (2005) 043505,
  [\href{http://xxx.lanl.gov/abs/hep-ph/0506320}{{\tt hep-ph/0506320}}].

\bibitem{Farrar:2005zd}
G.~R. Farrar and G.~Zaharijas, {\it {Dark matter and the baryon asymmetry}},
  {\em Phys.Rev.Lett.} {\bf 96} (2006) 041302,
  [\href{http://xxx.lanl.gov/abs/hep-ph/0510079}{{\tt hep-ph/0510079}}].

\bibitem{Suematsu:2005kp}
D.~Suematsu, {\it {Nonthermal production of baryon and dark matter}},  {\em
  Astropart.Phys.} {\bf 24} (2006) 511--519,
  [\href{http://xxx.lanl.gov/abs/hep-ph/0510251}{{\tt hep-ph/0510251}}].

\bibitem{Tytgat:2006wy}
M.~H. Tytgat, {\it {Relating leptogenesis and dark matter}},
  \href{http://xxx.lanl.gov/abs/hep-ph/0606140}{{\tt hep-ph/0606140}}.

\bibitem{Banks:2006xr}
T.~Banks, S.~Echols, and J.~Jones, {\it {Baryogenesis, dark matter and the
  Pentagon}},  {\em JHEP} {\bf 0611} (2006) 046,
  [\href{http://xxx.lanl.gov/abs/hep-ph/0608104}{{\tt hep-ph/0608104}}].

\bibitem{Khlopov:2008ty}
M.~Y. Khlopov and C.~Kouvaris, {\it {Composite dark matter from a model with
  composite Higgs boson}},  {\em Phys. Rev.} {\bf D78} (2008) 065040,
  [\href{http://xxx.lanl.gov/abs/0806.1191}{{\tt arXiv:0806.1191}}].

\bibitem{Kitano:2008tk}
R.~Kitano, H.~Murayama, and M.~Ratz, {\it {Unified origin of baryons and dark
  matter}},  {\em Phys.Lett.} {\bf B669} (2008) 145--149,
  [\href{http://xxx.lanl.gov/abs/0807.4313}{{\tt arXiv:0807.4313}}].

\bibitem{Kaplan:2009ag}
D.~E. Kaplan, M.~A. Luty, and K.~M. Zurek, {\it {Asymmetric Dark Matter}},
  {\em Phys.Rev.} {\bf D79} (2009) 115016,
  [\href{http://xxx.lanl.gov/abs/0901.4117}{{\tt arXiv:0901.4117}}].

\bibitem{Kohri:2009yn}
K.~Kohri, A.~Mazumdar, N.~Sahu, and P.~Stephens, {\it {Probing Unified Origin
  of Dark Matter and Baryon Asymmetry at PAMELA/Fermi}},  {\em Phys. Rev.} {\bf
  D80} (2009) 061302, [\href{http://xxx.lanl.gov/abs/0907.0622}{{\tt
  arXiv:0907.0622}}].

\bibitem{Kribs:2009fy}
G.~D. Kribs, T.~S. Roy, J.~Terning, and K.~M. Zurek, {\it {Quirky Composite
  Dark Matter}},  {\em Phys.Rev.} {\bf D81} (2010) 095001,
  [\href{http://xxx.lanl.gov/abs/0909.2034}{{\tt arXiv:0909.2034}}].

\bibitem{Cohen:2009fz}
T.~Cohen and K.~M. Zurek, {\it {Leptophilic Dark Matter from the Lepton
  Asymmetry}},  {\em Phys.Rev.Lett.} {\bf 104} (2010) 101301,
  [\href{http://xxx.lanl.gov/abs/0909.2035}{{\tt arXiv:0909.2035}}].

\bibitem{An:2009vq}
H.~An, S.-L. Chen, R.~N. Mohapatra, and Y.~Zhang, {\it {Leptogenesis as a
  Common Origin for Matter and Dark Matter}},  {\em JHEP} {\bf 1003} (2010)
  124, [\href{http://xxx.lanl.gov/abs/0911.4463}{{\tt arXiv:0911.4463}}].

\bibitem{Khlopov:2010pq}
M.~Y. Khlopov, A.~G. Mayorov, and E.~Y. Soldatov, {\it {Composite Dark Matter
  and Puzzles of Dark Matter Searches}},  {\em Int. J. Mod. Phys.} {\bf D19}
  (2010) 1385--1395, [\href{http://xxx.lanl.gov/abs/1003.1144}{{\tt
  arXiv:1003.1144}}].

\bibitem{Cohen:2010kn}
T.~Cohen, D.~J. Phalen, A.~Pierce, and K.~M. Zurek, {\it {Asymmetric Dark
  Matter from a GeV Hidden Sector}},  {\em Phys.Rev.} {\bf D82} (2010) 056001,
  [\href{http://xxx.lanl.gov/abs/1005.1655}{{\tt arXiv:1005.1655}}].

\bibitem{Shelton:2010ta}
J.~Shelton and K.~M. Zurek, {\it {Darkogenesis: A baryon asymmetry from the
  dark matter sector}},  {\em Phys.Rev.} {\bf D82} (2010) 123512,
  [\href{http://xxx.lanl.gov/abs/1008.1997}{{\tt arXiv:1008.1997}}].

\bibitem{Davoudiasl:2010am}
H.~Davoudiasl, D.~E. Morrissey, K.~Sigurdson, and S.~Tulin, {\it {Hylogenesis:
  A Unified Origin for Baryonic Visible Matter and Antibaryonic Dark Matter}},
  {\em Phys.Rev.Lett.} {\bf 105} (2010) 211304,
  [\href{http://xxx.lanl.gov/abs/1008.2399}{{\tt arXiv:1008.2399}}].

\bibitem{Haba:2010bm}
N.~Haba and S.~Matsumoto, {\it {Baryogenesis from Dark Sector}},
  \href{http://xxx.lanl.gov/abs/1008.2487}{{\tt arXiv:1008.2487}}.

\bibitem{Buckley:2010ui}
M.~R. Buckley and L.~Randall, {\it {Xogenesis}},
  \href{http://xxx.lanl.gov/abs/1009.0270}{{\tt arXiv:1009.0270}}.

\bibitem{Gu:2010ft}
P.-H. Gu, M.~Lindner, U.~Sarkar, and X.~Zhang, {\it {WIMP Dark Matter and
  Baryogenesis}},  \href{http://xxx.lanl.gov/abs/1009.2690}{{\tt
  arXiv:1009.2690}}.

\bibitem{Blennow:2010qp}
M.~Blennow, B.~Dasgupta, E.~Fernandez-Martinez, and N.~Rius, {\it {Aidnogenesis
  via Leptogenesis and Dark Sphalerons}},  {\em JHEP} {\bf 1103} (2011) 014,
  [\href{http://xxx.lanl.gov/abs/1009.3159}{{\tt arXiv:1009.3159}}].

\bibitem{Hall:2010jx}
L.~J. Hall, J.~March-Russell, and S.~M. West, {\it {A Unified Theory of Matter
  Genesis: Asymmetric Freeze-In}},
  \href{http://xxx.lanl.gov/abs/1010.0245}{{\tt arXiv:1010.0245}}.

\bibitem{Dutta:2010va}
B.~Dutta and J.~Kumar, {\it {Asymmetric Dark Matter from Hidden Sector
  Baryogenesis}},  \href{http://xxx.lanl.gov/abs/1012.1341}{{\tt
  arXiv:1012.1341}}.

\bibitem{Falkowski:2011xh}
A.~Falkowski, J.~T. Ruderman, and T.~Volansky, {\it {Asymmetric Dark Matter
  from Leptogenesis}},  \href{http://xxx.lanl.gov/abs/1101.4936}{{\tt
  arXiv:1101.4936}}.

\bibitem{Heckman:2011sw}
J.~J. Heckman and S.-J. Rey, {\it {Baryon and Dark Matter Genesis from Strongly
  Coupled Strings}},  \href{http://xxx.lanl.gov/abs/1102.5346}{{\tt
  arXiv:1102.5346}}.

\bibitem{Graesser:2011wi}
M.~L. Graesser, I.~M. Shoemaker, and L.~Vecchi, {\it {Asymmetric WIMP dark
  matter}},  \href{http://xxx.lanl.gov/abs/1103.2771}{{\tt arXiv:1103.2771}}.

\bibitem{Frandsen:2011kt}
M.~T. Frandsen, S.~Sarkar, and K.~Schmidt-Hoberg, {\it {Light asymmetric dark
  matter from new strong dynamics}},
  \href{http://xxx.lanl.gov/abs/1103.4350}{{\tt arXiv:1103.4350}}. * Temporary
  entry *.

\bibitem{Buckley:2011kk}
M.~R. Buckley, {\it {Asymmetric Dark Matter and Effective Operators}},
  \href{http://xxx.lanl.gov/abs/1104.1429}{{\tt arXiv:1104.1429}}.

\bibitem{Hook:2011tk}
A.~Hook, {\it {Unitarity constraints on asymmetric freeze-in}},
  \href{http://xxx.lanl.gov/abs/1105.3728}{{\tt arXiv:1105.3728}}.

\bibitem{Cheung:2011if}
C.~Cheung and K.~M. Zurek, {\it {Affleck-Dine Cogenesis}},
  \href{http://xxx.lanl.gov/abs/1105.4612}{{\tt arXiv:1105.4612}}.

\bibitem{DelNobile:2011je}
E.~Del~Nobile, C.~Kouvaris, and F.~Sannino, {\it {Interfering Composite
  Asymmetric Dark Matter for DAMA and CoGeNT}},
  \href{http://xxx.lanl.gov/abs/1105.5431}{{\tt arXiv:1105.5431}}.

\bibitem{Sakharov:1967dj}
A.~Sakharov, {\it {Violation of CP Invariance, c Asymmetry, and Baryon
  Asymmetry of the Universe}},  {\em Pisma Zh.Eksp.Teor.Fiz.} {\bf 5} (1967)
  32--35. Reprinted in *Kolb, E.W. (ed.), Turner, M.S. (ed.): The early
  universe* 371-373, and in *Lindley, D. (ed.) et al.: Cosmology and particle
  physics* 106-109, and in Sov. Phys. Usp. 34 (1991) 392-393 [Usp. Fiz. Nauk
  161 (1991) No. 5 61-64].

\bibitem{Cohen:1987vi}
A.~G. Cohen and D.~B. Kaplan, {\it {Thermodynamic generation of the baryon
  asymmetry}},  {\em Phys.Lett.} {\bf B199} (1987) 251.

\bibitem{Cohen:1988kt}
A.~G. Cohen and D.~B. Kaplan, {\it {Spontaneous baryogenesis}},  {\em
  Nucl.Phys.} {\bf B308} (1988) 913.

\bibitem{Cohen:1991iu}
A.~G. Cohen, D.~Kaplan, and A.~Nelson, {\it {Spontaneous baryogenesis at the
  weak phase transition}},  {\em Phys.Lett.} {\bf B263} (1991) 86--92.

\bibitem{Abel:1992za}
S.~Abel, W.~Cottingham, and I.~Whittingham, {\it {Spontaneous baryogenesis in
  supersymmetric models}},  {\em Nucl.Phys.} {\bf B410} (1993) 173--187,
  [\href{http://xxx.lanl.gov/abs/hep-ph/9212299}{{\tt hep-ph/9212299}}].

\bibitem{Cohen:1994ss}
A.~G. Cohen, D.~Kaplan, and A.~Nelson, {\it {Diffusion enhances spontaneous
  electroweak baryogenesis}},  {\em Phys.Lett.} {\bf B336} (1994) 41--47,
  [\href{http://xxx.lanl.gov/abs/hep-ph/9406345}{{\tt hep-ph/9406345}}].

\bibitem{Dolgov:1994zq}
A.~Dolgov and K.~Freese, {\it {Calculation of particle production by Nambu
  Goldstone bosons with application to inflation reheating and baryogenesis}},
  {\em Phys.Rev.} {\bf D51} (1995) 2693--2702,
  [\href{http://xxx.lanl.gov/abs/hep-ph/9410346}{{\tt hep-ph/9410346}}].

\bibitem{Dolgov:1996qq}
A.~Dolgov, K.~Freese, R.~Rangarajan, and M.~Srednicki, {\it {Baryogenesis
  during reheating in natural inflation and comments on spontaneous
  baryogenesis}},  {\em Phys.Rev.} {\bf D56} (1997) 6155--6165,
  [\href{http://xxx.lanl.gov/abs/hep-ph/9610405}{{\tt hep-ph/9610405}}].

\bibitem{Bertolami:1996cq}
O.~Bertolami, D.~Colladay, V.~Kostelecky, and R.~Potting, {\it {CPT violation
  and baryogenesis}},  {\em Phys.Lett.} {\bf B395} (1997) 178--183,
  [\href{http://xxx.lanl.gov/abs/hep-ph/9612437}{{\tt hep-ph/9612437}}].

\bibitem{Li:2001st}
M.-z. Li, X.-l. Wang, B.~Feng, and X.-m. Zhang, {\it {Quintessence and
  spontaneous leptogenesis}},  {\em Phys.Rev.} {\bf D65} (2002) 103511,
  [\href{http://xxx.lanl.gov/abs/hep-ph/0112069}{{\tt hep-ph/0112069}}].

\bibitem{Yamaguchi:2002vw}
M.~Yamaguchi, {\it {Generation of cosmological large lepton asymmetry from a
  rolling scalar field}},  {\em Phys.Rev.} {\bf D68} (2003) 063507,
  [\href{http://xxx.lanl.gov/abs/hep-ph/0211163}{{\tt hep-ph/0211163}}].

\bibitem{Brandenberger:2003kc}
R.~H. Brandenberger and M.~Yamaguchi, {\it {Spontaneous baryogenesis in warm
  inflation}},  {\em Phys.Rev.} {\bf D68} (2003) 023505,
  [\href{http://xxx.lanl.gov/abs/hep-ph/0301270}{{\tt hep-ph/0301270}}].

\bibitem{Takahashi:2003db}
F.~Takahashi and M.~Yamaguchi, {\it {Spontaneous baryogenesis in flat
  directions}},  {\em Phys.Rev.} {\bf D69} (2004) 083506,
  [\href{http://xxx.lanl.gov/abs/hep-ph/0308173}{{\tt hep-ph/0308173}}].

\bibitem{Alberghi:2003ws}
G.~Alberghi, R.~Casadio, and A.~Tronconi, {\it {Radion induced spontaneous
  baryogenesis}},  {\em Mod.Phys.Lett.} {\bf A22} (2007) 339--346,
  [\href{http://xxx.lanl.gov/abs/hep-ph/0310052}{{\tt hep-ph/0310052}}].

\bibitem{Carmona:2004xc}
J.~M. Carmona, J.~L. Cortes, A.~K. Das, J.~Gamboa, and F.~Mendez, {\it
  {Matter-antimatter asymmetry without departure from thermal equilibrium}},
  {\em Mod.Phys.Lett.} {\bf A21} (2006) 883--892,
  [\href{http://xxx.lanl.gov/abs/hep-th/0410143}{{\tt hep-th/0410143}}].

\bibitem{Carroll:2005dj}
S.~M. Carroll and J.~Shu, {\it {Models of baryogenesis via spontaneous Lorentz
  violation}},  {\em Phys.Rev.} {\bf D73} (2006) 103515,
  [\href{http://xxx.lanl.gov/abs/hep-ph/0510081}{{\tt hep-ph/0510081}}].

\bibitem{Barenboim:2007tu}
G.~Barenboim and J.~D. Lykken, {\it {Quintessence, inflation and baryogenesis
  from a single pseudo-Nambu-Goldstone boson}},  {\em JHEP} {\bf 0710} (2007)
  032, [\href{http://xxx.lanl.gov/abs/0707.3999}{{\tt arXiv:0707.3999}}].

\bibitem{Harvey:1990qw}
J.~A. Harvey and M.~S. Turner, {\it {Cosmological baryon and lepton number in
  the presence of electroweak fermion number violation}},  {\em Phys.Rev.} {\bf
  D42} (1990) 3344--3349.

\bibitem{Chung:2008gv}
D.~J.~H. Chung, B.~Garbrecht, and S.~Tulin, {\it {The Effect of the Sparticle
  Mass Spectrum on the Conversion of B-L to B}},  {\em JCAP} {\bf 0903} (2009)
  008, [\href{http://xxx.lanl.gov/abs/0807.2283}{{\tt arXiv:0807.2283}}].

\bibitem{Komatsu:2010fb}
{\bf WMAP} Collaboration, E.~Komatsu {\em et.~al.}, {\it {Seven-Year Wilkinson
  Microwave Anisotropy Probe (WMAP) Observations: Cosmological
  Interpretation}},  {\em Astrophys. J. Suppl.} {\bf 192} (2011) 18,
  [\href{http://xxx.lanl.gov/abs/1001.4538}{{\tt arXiv:1001.4538}}].

\bibitem{Svrcek:2006yi}
P.~Svrcek and E.~Witten, {\it {Axions In String Theory}},  {\em JHEP} {\bf
  0606} (2006) 051, [\href{http://xxx.lanl.gov/abs/hep-th/0605206}{{\tt
  hep-th/0605206}}].

\bibitem{Arvanitaki:2009fg}
A.~Arvanitaki, S.~Dimopoulos, S.~Dubovsky, N.~Kaloper, and J.~March-Russell,
  {\it {String Axiverse}},  {\em Phys.Rev.} {\bf D81} (2010) 123530,
  [\href{http://xxx.lanl.gov/abs/0905.4720}{{\tt arXiv:0905.4720}}].

\bibitem{Rajaraman:2011wf}
A.~Rajaraman, W.~Shepherd, T.~M. Tait, and A.~M. Wijangco, {\it {LHC Bounds on
  Interactions of Dark Matter}},  \href{http://xxx.lanl.gov/abs/1108.1196}{{\tt
  arXiv:1108.1196}}.

\bibitem{Buckley:2009in}
M.~R. Buckley and P.~J. Fox, {\it {Dark Matter Self-Interactions and Light
  Force Carriers}},  {\em Phys.Rev.} {\bf D81} (2010) 083522,
  [\href{http://xxx.lanl.gov/abs/0911.3898}{{\tt arXiv:0911.3898}}].

\end{thebibliography}\endgroup

\end{document}